# 旋转抛物面球形接收器系统聚光性能分析与优化

黄卫东 [1]，黄法融 [1]，胡芃 [2]，陈则韶 [2]
中国科学技术大学 [1]地球与空间科学学院，[2]工程科学学院
安徽合肥金寨路 96 号，230026， email: huangwd@ustc.edu.cn

摘要：旋转抛物面球形接收器是较少受到关注的太阳能聚光系统，本文发展了真实太阳模型和高斯太阳模型下，计算带球形接收器的旋转抛物面碟式聚光系统拦截率和效率的表达式，给出了计算参数，可以快速准确计算系统效率。我们首先计算接收器对抛物面反射镜上任一点所反射的光线的拦截率，再通过对整个镜面积分计算系统拦截率和光学效率。我们比较了两种太阳模型的计算结果，给出了高斯太阳模型的适用范围。更进一步，我们提出了优化系统性能的快速方法，计算了典型设计条件下的优化效率与光学误差的关系，结果表明，旋转抛物面球形接收器碟式系统的热效率要比塔式和槽式系统高 20%以上，主要体现在热损小，余弦因子高，拦截率高等方面。

关键词：碟式，球面接收器，高斯分布，真实太阳，效率，拦截率

0、引言

　　旋转抛物面将平行光反射到焦点上，可以在焦点上获得较高温度，是太阳能热利用的主要方法之一(Mancini, Heller et al. 2003)。当使用旋转抛物面聚焦太阳光时，太阳光线不是严格的平行光，而且反射镜及其跟踪系统的光学误差，常常扩大了反射光线到焦点的方向偏差(Shuai, Xia et al. 2008)。人们需要了解反射光线的空间分布，计算接收器对反射光线的拦截率，从而获得聚光系统的光学性能。

　　传统上，人们利用光线追踪法，由计算机模拟太阳上不同位置的光线经反射镜反射的方向，获得接收器对反射光线的拦截率(Jiang, Hu et al. 2010)。这种方法可以详细模拟各种光学过程，考虑反射镜材料，形状误差，跟踪误差，接收器形状，材料性质和安装误差等一系列因素对系统性能的影响，但是使用这种方法，无法了解系统性能和各种因素之间的函数关系(Bendt and Rabl 1981)。

　　另外一类计算系统光学性能的方法是先计算接收器上能流密度分布，然后积分接收器上能流密度分布得到系统收集的总能量(Jeter 1987)。人们发展了多种积分表达式计算接收器上能流密度分布(Kamada 1965; Harris and Duff 1981; HONG and LEE 1987; Jones and Wang 1995; Elsayed and Fathalah 1996; Ruiheng, Yuming et al. 2005; Liu, Dai et al. 2007)，由于这些计算能流密度分布的被积函数就非常复杂，难以得到积分的解析解，只能进行数值积分，从而无法了解系统性能与各种因素之间的函数关系。

　　O'neill 和 Hudson 最早为安装平板接收器的抛物面碟式系统提出了解析表达式(O'Neill and Hudson 1978)。该方法能够快速计算光学性能，但是，方法是建立在均匀太阳光球模型基础上的，与实际太阳光强分布存在明显差别,因为太阳边缘存在昏暗效应(Negi, Bhowmik et al. 1986)。Bendt and Rabl 则通过角接受函数，以反射光强高斯分布为基础，建立了能够快速计算带球形或平板接收器的抛物面碟式系统光学性能的函数(Bendt and Rabl 1981)。Stine and Harrigan 同样基于高斯分布假设，提出了另外一种计算碟式系统拦截率函数(Stine and Harrigan 1985)。但是，当光学误差较小时，根据高斯分布计算得到的拦截率会相差 10%以上(Bendt, Rabi et al. 1979)。因此，以实际太阳光强分布模型为基础，建立计算碟式系统性能的方法，是非常必要的。

　　本文主要针对旋转抛物面球形接收器聚光系统。通常球形接收器将光能转化为高温热能

(Kamada 1965)，通过介质传输，用于发电和工业，或直接生产氢气。目前在文献中报道较少。我们提出了真实太阳亮度分布和高斯分布近似下，计算该聚光反射系统拦截率的方法。我们首先计算每个反射点反射的光线的拦截率，再对整个反射面进行积分，计算系统平均拦截率，从理论上推出了计算函数。在此基础上，我们给出了优化球形接收器碟式系统效率的方法和计算结果。

1、球形接收器对旋转抛物面反射光线的拦截率

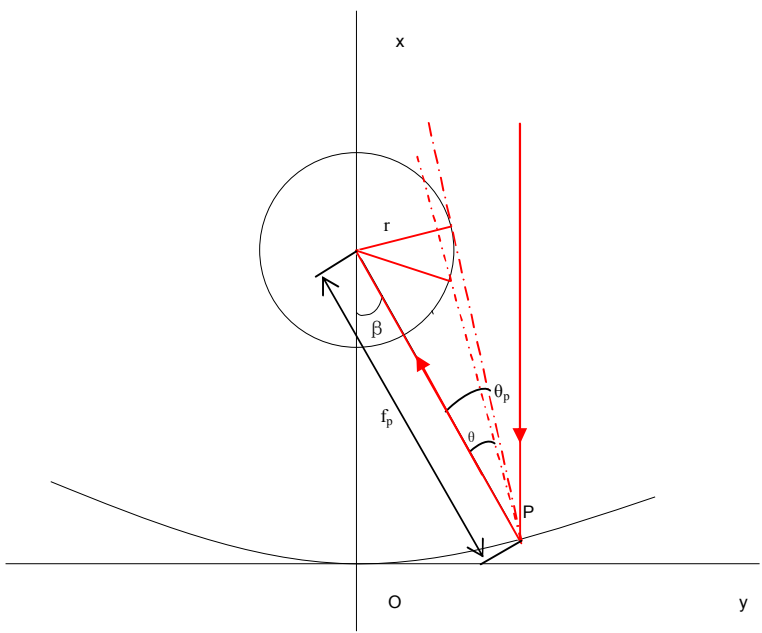

图 2 旋转抛物面球形接收器碟式系统光学效率计算示意图

如图 2 是安装球形接收器的碟式聚光系统横截面图，球形接收器的中心在旋转抛物面焦点上。抛物面顶点是坐标原点，对称轴是 x 轴，对镜面任意一点 P，过反射点 P 所在面是 xoy 面。θ 是太阳中心光线与任意一条光线经反射后的反射线之间的夹角。假设接收器对 P 的半张角为 $\theta_P$，也就是说，从 P 点反射，与接收器相切的反射光线与太阳中心光线的反射线之间夹角为 $\theta_P$，$\theta_P$ 可按下式计算

$$\theta_P = \mathrm{asin}(r/f_P) \tag{1}$$

r 是接收器半径，$f_P$ 是 P 点到焦点距离，可按下式计算。

$$f_P = f/\cos^2(\beta/2) = f + y^2/(4f) \tag{2}$$

这里 β 是 P 点边缘角，y 是 P 点到抛物面对称轴距离，f 是抛物面反射镜焦距。当来自太阳的光线被 P 点反射，考虑反射光线中夹角为 θ 与 θ+dθ 之间的一束环形光线，则当 θ<=$\theta_P$ 时，该束光线被接收器拦截，光线强度为 B(θ)*2πθdθ I*dS，I 是入射太阳光强，dS 是 P 点附近区域面积，所以该点反射光线被拦截的比例为：

$$\gamma_P(y) = \int_0^{\theta_P} B_{eff}(\theta) * 2\pi\theta d\theta IdS/(IdS) = \int_0^{\theta_P} B_{eff}(\theta) * 2\pi\theta d\theta \tag{3}$$

这里 $B_{eff}$(θ) 是归一化的反射光强分布函数。

得到反射镜上每个反射点的光线效率后，我们可以积分计算整个反射镜被拦截的总光强，其表达式如下：

$$I = \int_r^R \gamma_P(y) * I_{in} * 2\pi y dy + \pi r^2 I_{in} \tag{4}$$

$I_{in}$ 是单位面积上反射光线强度，r 是接收器外半径，R 是反射镜半径。该式前一个积分是反射镜反射光线贡献，后一个是太阳光直接照射到接收器上的贡献，积分上限从 r 开始，是考虑了接受器的遮盖作用。

所以系统对入射光线的拦截率则可按下式计算：

$$\gamma = \frac{\int_r^R \gamma_P(y) * I_{in} * 2\pi y dy + \pi r^2 I_{in}}{\pi R^2 I_{in}} = \frac{\int_r^R \gamma_P(y) * 2y dy}{R^2} + (\frac{r}{R})^2 \tag{5}$$

将（3）式带入（5）式，则我们得到：

$$\gamma = \frac{\int_r^R \int_0^{\theta_P} B(\theta) 2\pi\theta d\theta * 2y dy}{R^2} + (\frac{r}{R})^2 \tag{6}$$

由于接收器对反射镜上任意一点的张角 $\theta_P$ 都很小，为了化简上式，我们引入近似：

$\theta_P =$ asin（r/f$_P$）≈r/f$_P$ =r/f*cos$^2$(β/2)    (7)

将其引入到（1）和（2）式，我们得到：

2ydy＝－4f$^2$θ$_0$dθ$_P$/θ$_P^2$    (8)

这里 θ$_0$＝asin(r/f)，是接收器对反射镜顶点的张角。将（8）式带入（6）式，则我们得到：

$$\gamma = \frac{\int_{\theta_r}^{\theta_R} \int_0^{\theta_P} -B(\theta) 2\pi\theta d\theta * 4f_0^2 \theta_0 / \theta_P^2 d\theta_P}{R^2} + (\frac{r}{R})^2 \tag{9}$$

$$= \frac{\int_{\theta_r}^{\theta_R} [-\int_0^{\theta_P} B(\theta) 2\pi\theta d\theta] * \theta_0 / \theta_P^2 d\theta_P}{\tan^2(\beta_R/2)} + (\frac{\tan(\beta_r/2)}{\tan(\beta_R/2)})^2 \tag{10}$$

这里 θ$_r$ 和 θ$_R$ 分别对应 y＝r 和 y＝R 时，接收器的张角。

下面我们分析，当引入适当形式的反射光强分布表达式时，可以得到上述积分的解析解，从而得到表达旋转抛物面球形接收器聚光系统光学性能的计算式。

2、高斯分布近似下以球面为接收器的碟式系统光学性能

如果反射光强分布是一维高斯分布，计算就会大大简化。当系统光学误差和光源都是高斯分布时，反射光强就是高斯分布。如果非高斯部分的分布宽度比高斯分布宽度小得多，这时，将反射光强分布当做高斯分布是很好的近似(Butler and Pettit 1977; Bendt and Rabl 1981)。例如，晴朗天气下，太阳光强分布宽度是 2.6mrad(Bendt and Rabl 1979)，虽然太阳光强分布不是高斯分布，当系统光学误差分布是高斯分布，而且分布宽度较大时，我们仍然可以将反射光强分布近似为高斯分布，分布函数为：

$$B_{eff,Gauss}(\theta) = \frac{1}{2\pi\sigma_t^2} \exp(-\frac{\theta^2}{2\sigma_t^2}) \tag{11}$$

分布宽度可按下式计算：

$$\sigma_t^2 = \sigma_{optic}^2 + \sigma_{sun}^2 \tag{12}$$

Biggs 和 Vittoe 的结果表明，引入高斯近似，在光学误差大于 10mrad 时，带来的误差小于 1%(Biggs and Vittoe 1979)。

将（11）式代入（3）式得到：

$$\gamma_P = 1 - \exp(-\frac{\theta_P^2}{2\sigma_t^2}) \tag{13}$$

将其代入（10）式，可得到：

$$\gamma = \frac{\int_{\theta_r}^{\theta_R} -[1-\exp(-\frac{\theta_P^2}{2\sigma_t^2})]*4f_0^2\theta_0/\theta_P^2 d\theta_P}{R^2} + (\frac{r}{R})^2 \tag{14}$$

$$= \frac{\int_{\theta_r}^{\theta_R} -[1-\exp(-\frac{\theta_P^2}{2\sigma_t^2})]*\theta_0/\theta_P^2 d\theta_P}{\tan^2(\beta_R/2)} + (\frac{\tan(\beta_r/2)}{\tan(\beta_R/2)})^2 \tag{15}$$

令 k＝θ$_0$/σ$_t$；考虑接收器很小，忽略接收器的遮盖效应时，上式可简化为：

$$\gamma = 1 + \frac{\exp(-\frac{k^2}{2})}{\tan^2\frac{\beta_R}{2}} - \frac{\exp(-\frac{k^2}{2}\cos^4\frac{\beta_R}{2})}{\sin^2\frac{\beta_R}{2}} - \frac{k}{\tan^2\frac{\beta_R}{2}}\int_k^{k\cos^2(\beta_R/2)}\exp(-\frac{t^2}{2})dt$$

$$= 1 + \frac{\exp(-\frac{k^2}{2})}{\tan^2\frac{\beta_R}{2}} - \frac{\exp(-\frac{k^2}{2}\cos^4\frac{\beta_R}{2})}{\sin^2\frac{\beta_R}{2}} - \frac{\sqrt{\pi}/2*k}{\tan^2\frac{\beta_R}{2}}[erf(\frac{k}{\sqrt{2}}\cos^2\frac{\beta_R}{2}) - erf(\frac{k}{\sqrt{2}})] \tag{16}$$

erf 是误差函数。本文得到的积分式 15 虽然与 Bendt and Rabl 形式不同(Bendt and Rabl 1981)，但是，Bendt and Rabl 的结果同样可以积分成式 16，两种方法所得到的结果是一致的。从上式可以看出，在高斯分布近似下，系统平均拦截率主要与边缘角和 k＝(r/f)/σ$_t$ 相关。r/(fσ$_t$)是接收器对反射镜顶点张角与反射光强高斯分布宽度之比.

我们根据实际太阳光强分布和聚光光学系统光学误差计算反射光强分布，用最小二乘拟合法，获得反射光强高斯分布的参数 σ$_t$，从而使反射光强高斯分布与实际反射光强分布的误差最小，获得高斯分布近似下最佳结果。

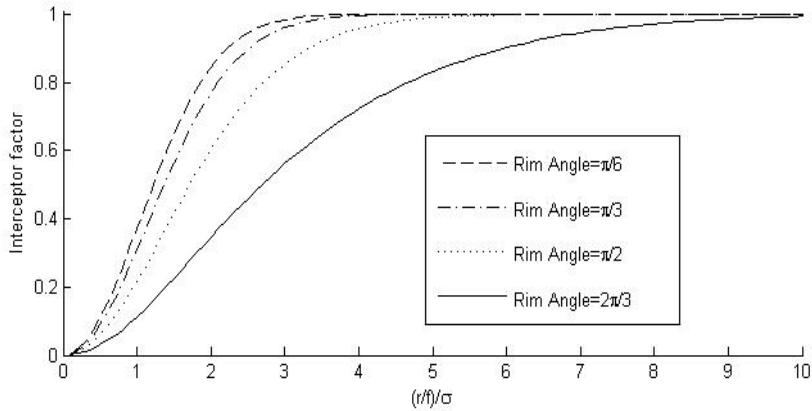

图 3 不同边缘角下，旋转抛物面球形接收器系统拦截率对接收器张角与反射光强分布宽度之比的关系

图 3 是根据计算式 16 计算得到的 4 种边缘角下，系统拦截率与接收器张角与反射光强分布宽度之比的关系，从图中可以看出，系统拦截率随比值 k=(r/f)/σ$_t$ 的增大而增大，在比值达到 6－10 以后，拦截率接近 100%。

根据我们提出的计算反射镜上某点的反射光线的拦截率（式 13），它只与该反射点 k=(r/f$_P$)/σ$_t$ 相关，没有被拦截部分随 k=(r/f$_P$)/σ$_t$ 的平方成指数下降，在 k 为 3 时，拦截率接近 99%。另一方面，k=(r/f$_P$)/σ$_t$ 与反射点所在边缘角一半的余弦的平方成正比，随着边缘角增大，k 值减小，从而降低了拦截率。系统拦截率是对所有反射点拦截率的平均，由于边缘角较大的部分占据较大比例，从而使系统拦截率随边缘角增大而明显下降。

3、实际太阳光强分布下系统拦截率

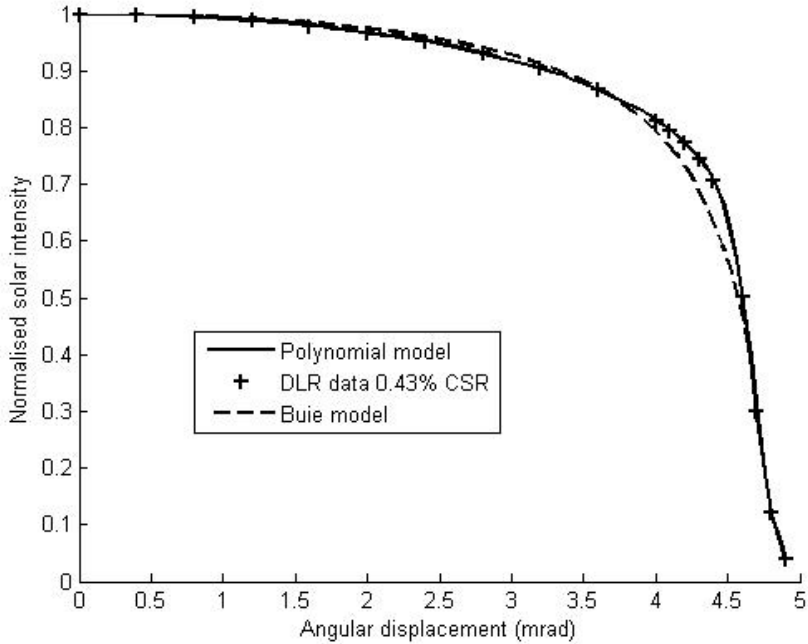

图 4 太阳光强径向分布

在光学误差较大，大于太阳宽度时，高斯近似能够给出误差小于 1% 的结果，但是，当光学误差较小，小于或与太阳宽度相当时，由于太阳光强分布与高斯分布偏离较大，高斯近似带来的误差会大于 10%(Bendt, Rabi et al. 1979)，因此，考虑实际太阳光强分布是非常必要的。

太阳光强分布函数常表示为径向亮度分布 B$_{radial}$(θ)，单位是 W/m$^2$sr，sr 是球面度，θ 是从太阳中心测量的角度。描述太阳光强分布的模型很多，例如，最近 Buie 根据太阳光强分布的测量数据给出了通用分析模型(Buie, Monger et al. 2003)，但是，该模型仍然有一定误差，我们根据实测资料(Neumann, Witzke et al. 2002)，使用多项式拟合模型描述太阳光球和部分光晕的辐射能量分布，如下式所示:

$$B_{sun}(\theta) = \sum_{i=0}^{N} a_i \theta^i \quad (17)$$

其拟合效果参见图 2，同时使用 Burie 指数下降模型描述其余光晕(Buie, Monger et al. 2003)。从图中可以看到，采用多项式模拟可以最大限度拟合实际太阳光球和光球附近的光强分布。

在实际聚光器上存在光学误差，如镜面形状，位置和倾角误差，接收器位置偏差，跟踪

偏差等。Bendt 等根据时间和大面积反射镜面平均，这些误差可近似看成高斯分布(Bendt, Rabi et al. 1979)，将光学误差分布函数与上式卷积，可得到反射光强分布函数 $B_{eff}$（θ）：

$$B_{eff}(\theta) = \frac{1}{\sigma_{optic}^2} \iint d^2\theta' \exp(-\frac{\theta'^2}{2\sigma_{optic}^2}) B_{Sun}(\theta - \theta') \tag{18}$$

为了加快计算速度，我们采用多项式拟合上式计算得到的反射光强分布：

$$B_{eff}(\theta) = \sum_{i=0}^{N} b_i \theta^{2i} \tag{19}$$

N＝10 时，计算得到不同光学误差下，反射光强分布的参数 $b_i$ 见表 1

将上式代入到（3）式，可得到：

$$\gamma_P(\theta_P) = \pi \sum_{I=0}^{N} \frac{b_i \theta_P^{2i+2}}{i+1} \tag{20}$$

$$\begin{aligned}
\gamma &= \frac{\int_{\theta_r}^{\theta_R} -\pi \sum_{i=0}^{N} \frac{b_i \theta_P^{2i+2}}{i+1} * 4f_0^2 \theta_0 / \theta_P^2 d\theta_P}{R^2} + (\frac{r}{R})^2 \\
&\approx \frac{\pi \theta_0}{\tan^2(\beta_R/2)} \sum_{i=1}^{N} \frac{b_i \theta_P^{2i+1}}{(2i+1)*(i+1)} \bigg|_{\theta_R}^{\theta_r} + (\frac{\theta_0}{2\tan(\beta_R/2)})^2 \\
&\approx \frac{\pi \theta_0}{\tan^2(\beta_R/2)} \sum_{i=1}^{N} \frac{b_i \theta_P^{2i+1}}{(2i+1)*(i+1)} \bigg|_{\theta_R}^{\theta_0}
\end{aligned} \tag{21}$$

由上式可知，在真实太阳亮度分布下，系统平均拦截率与反射光强分布，接收器对反射镜张角及反射镜边缘角相关。

使用真实太阳模型计算球形接收器碟式系统拦截率，根据我们得到的计算式，主要与系统光学误差，边缘角和接收器张角相关。太阳光强分布和系统光学误差共同决定了反射光强分布。在光学误差较小时，使用高斯分布来拟合反射光强分布，误差较大。我们使用多项式拟合太阳光球及其附近区域的光强分布，这时得到的反射光强分布不是高斯分布，为了节省计算时间，我们使用多项式拟合反射光强分布，对于具有不同光学误差的碟式系统，拟合多项式参数不同，表 1 给出了光学误差小于 **5mrad** 下的拟合反射光强分布的多项式参数。

图 5 给出了五种光学误差，在四种边缘角下的碟式系统拦截率随 k＝(r/$f_P$)/$σ_t$ 的变化。在相同光学误差和边缘角下，系统拦截率随接收器张角增大而增大。在相同接收器张角下，边缘角越大，拦截率越低，光学误差越大，拦截率越低。

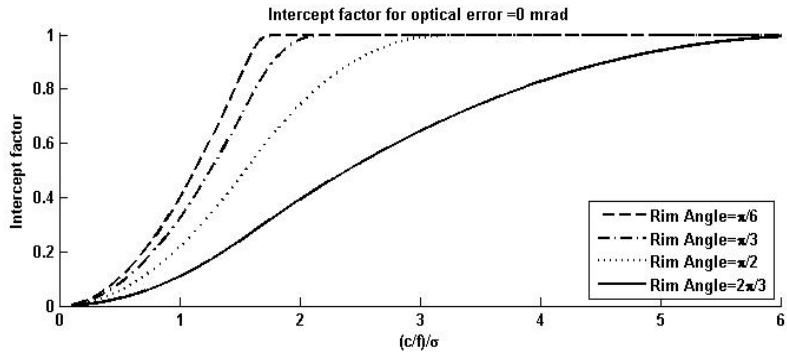
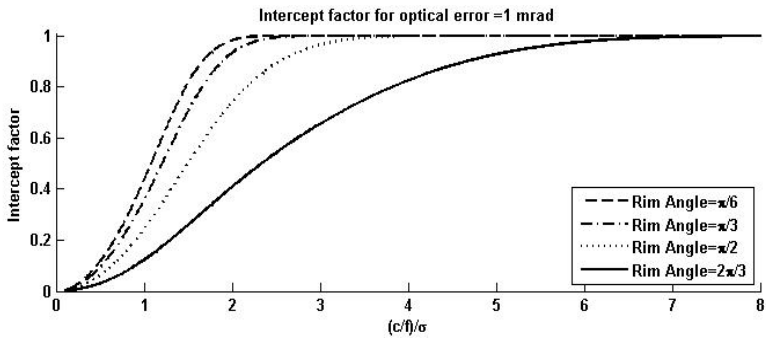
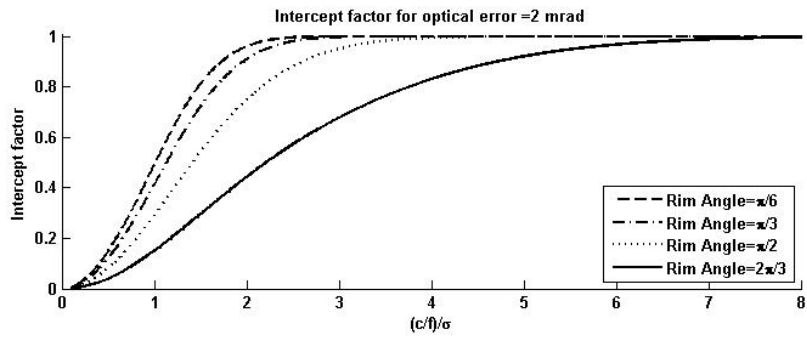
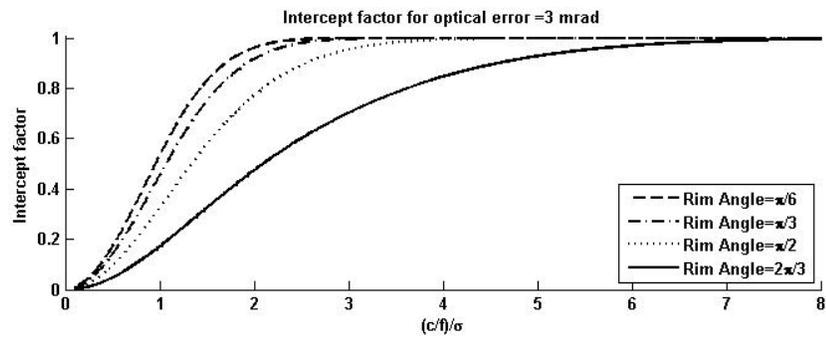

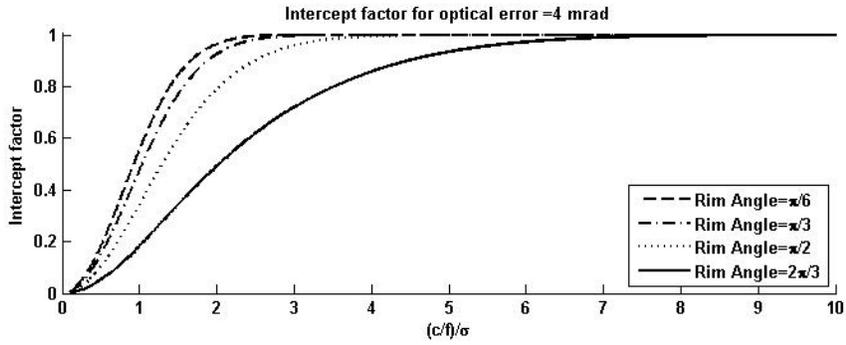

图 5 不同光学误差下，旋转抛物面球形接收器系统拦截率变化

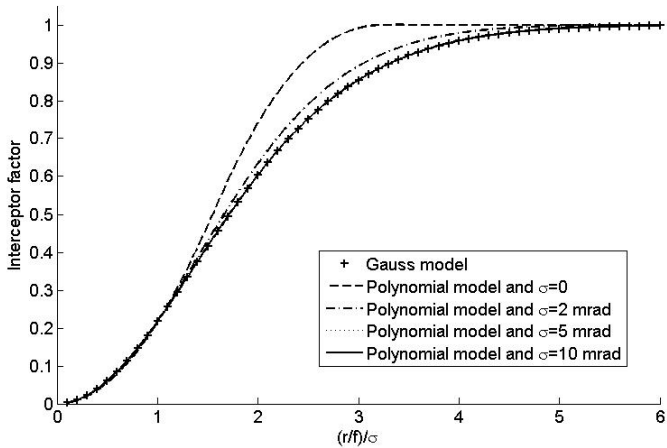

图 6 高斯模型和多项式拟合反射光强分布模型计算得到的旋转抛物面球形接收器系统拦截率
边缘角为 90 度

图 6 给出了两种太阳光强分布模型计算拦截率的差别。在 Gauss 模型中，拦截率只与 k ＝（r/f）/$\sigma_t$ 相关，而在真实太阳模型中，在光学误差较小时，拦截率不仅与 k 相关，而且与光学误差相关。但在光学误差较大时，拦截率只与 k 相关了，从而与高斯模型结果一致。在图中，就是光学误差为 5 和 10mrad 两种情况下，其拦截率曲线与高斯模型重合。图 7 给出了真实反射光强分布和反射光强高斯分布的对比。其中高斯分布是利用高斯分布函数拟合得到的，从图 7 中可以看出，在光学误差为 0 和 2mrad 情况下，高斯分布函数对实际反射光强分布的拟合误差较大，而在光学误差为 5 和 10mrad 情况下，用高斯分布拟合实际反射光强分布的误差很小。我们的计算表明，光学误差大于 5mrad 时，可以采用高斯分布函数来描述反射光强分布。

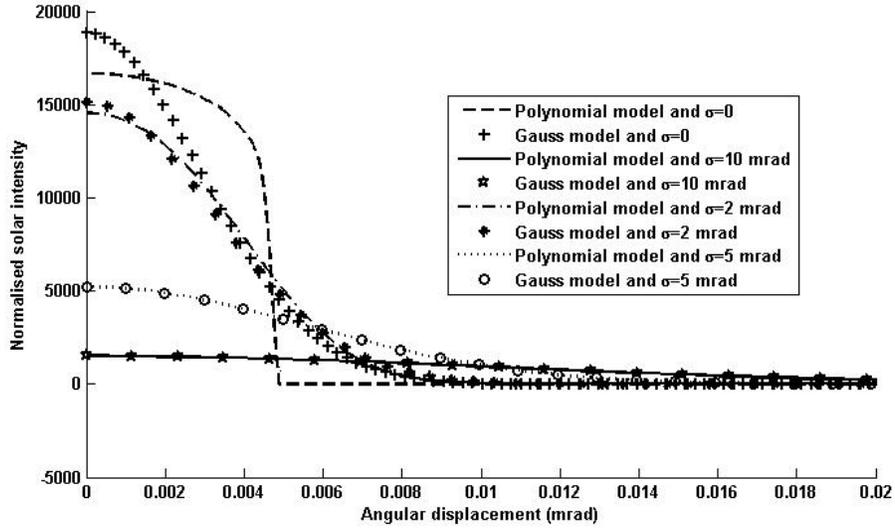

图 7 使用高斯分布拟合反射光强分布，σ 代表系统光学误差

4、系统优化

本文讨论使用球面接收器的抛物面碟式系统获得高温热能情况。为了简化问题，假定接收器工作在某个温度下，单位面积接收器上的热损 q 是不变的常数。例如，在工作温度为 800ºC 下，开口直径为 0.14m 的 WGA 空腔接收器的总热损功率是 0.260kW，随太阳高度角略有变化，在太阳高度角最大变化为 60 度时，变化约 2%(FRASER 2008)。下面我们使用该数据作为球形接收器热损，对系统进行优化。

系统优化的目标是使输出的热能最大化。在反射镜光学误差一定的情况下，系统在一年内输出的净热能（单位是 kWh）如下：

$$E = \pi R^2 \rho \alpha \gamma \int_0^{365} 24 n(t) DNI(t) dt - 4\pi r^2 q \int_0^{365} 24 n(t) g(t) dt \quad (22)$$

这里 ρ 是反射镜反射率，α 是接收器吸收率，n 代表 t 时刻实际直射辐射与天气情好时直射辐射之比，DNI（t）是系统收集能量时的太阳直射能量（kW/m²）（最好单位统一为 W/m2），为了简化计算，我们假定太阳高度角 h 大于等于 15 度时，系统开始收集能量，则根据晴天模型，我们有(Chen and Li 2003)：

$$DNI(t) = I_s(t) p^m \quad h>=15$$
$$= 0 \quad h<15 \quad (23)$$

$I_s(t)$是大气层外太阳辐射强度(W/m²)，$I_s(t)$=1367*(1+0.034*cos(2πt/365))，m 是大气质量，按下式计算：

$$m=[1229+(614\sinh)^2]^{1/2}-614\sinh \quad (24)$$

对于 g 来说，可以按下式计算：

$$g（t）=1 \quad h>=15$$
$$= 0 \quad h<15 \quad (25)$$

则上式可简化为：

$$E = \pi R^2 \rho \alpha \gamma E_0 - 4\pi r^2 q t_0 \quad (26)$$

这里 $E_0 = \int_0^{365} 24n\,DNI(t)\,dt$，$\int_0^{365} 24n(t)DNI(t)\,dt$ 代表了当地一年里太阳高度大于 15 度的晴天直射总能量（单位 kWh），而 $t_0 = \int_0^{365} 24nf(t)\,dt \int_0^{365} 24n(t)g(t)\,dt$ 是当地一年里太阳高度大于 15 度的晴天总时数（包括非晴天折算的晴天时数），对于特定地区来说，在优化系统时，它们与系数设计参数无关，可以看成是常数。

我们的目标之一是使系统单位通光面积收集的净能量 E1 最大，E1 计算式如下：

$$E1 = \frac{E}{\pi R^2} = \rho\alpha\gamma E_0 - 4r^2 q t_0 / R^2 \tag{27}$$

或者使系统净热效率或热效率（后面用的是这两个）η 最大，效率计算式如下：

$$\eta = \frac{E}{\pi R^2 E_0} = \rho\alpha\gamma - 4r^2 q t_0 / (E_0 R^2) \tag{28}$$

该式后一项是接收器热损能量与系统工作时接受的太阳直射能量之比。其中 $qt_0/E_0$ 是一年里单位接收器面积上热损能量与单位面积聚光器接受的太阳直射能量之比，我们称为系统能量损失系数 ξ，我们使用晴天模型近似来计算年平均 ξ，如下式：

$$\xi = \frac{qt_0}{E_0} = \frac{q\int_0^{365} 24n g(t)\,dt}{\int_0^{365} 24n\,DNI(t)\,dt} \approx \frac{q\int_0^{365} g(t)\,dt}{\int_0^{365} DNI(t)\,dt} \tag{29}$$

在下面分析中，我们将特定地区的碟式系统能量损失系数 ξ 视为常数。将式（21）和（29）代入到式（28），我们得到：

$$\eta = \rho\alpha \frac{\pi\theta_0}{\tan^2(\beta_R/2)} \sum_{i=1}^{N} \frac{b_i \theta_P^{2i+1}}{(2i+1)*(i+1)} \bigg|_{\theta_R}^{\theta_0} - \frac{\xi\theta_0^2}{\tan^2(\beta_R/2)} \tag{30}$$

在系统材料，光学误差和能量损失系数（ξ）确定的情况下，系统效率主要由接收器张角(open angle of receiver)$\theta_0$ 和旋转抛物面反射镜的边缘角 $\beta_R$ 决定。

当反射光强分布是高斯分布时，将式（16）和（29）代入式（28），我们得到：

$$\eta = \rho\alpha\left[1 + \frac{\exp(-\frac{k^2}{2})}{\tan^2\frac{\beta_R}{2}} - \frac{\exp(-\frac{k^2}{2}\cos^4\frac{\beta_R}{2})}{\sin^2\frac{\beta_R}{2}} - \frac{k}{\tan^2\frac{\beta_R}{2}} \int_k^{k\cos^2(\beta_R/2)} \exp(-\frac{t^2}{2})\,dt\right] - \frac{\xi\theta_0^2}{\tan^2(\beta_R/2)} \tag{31}$$

也就是说，在光学误差较大时，反射光强分布是高斯分布时，系统效率同样由接收器张角(open angle of receiver) $\theta_0$ 和旋转抛物面反射镜的边缘角 $\beta_R$ 决定。图 8 给出了一定设计条件下，包括光学误差为 5mrad，接收器能量损失系数 ξ＝20.674，反射率和吸收率系数乘积 ρα＝0.8784，系统热效率与边缘角和接收器张角变化关系。从图中可以看成，在很大的接收器张角范围和边缘角范围内，都有较高拦截率，高效率区范围较大。

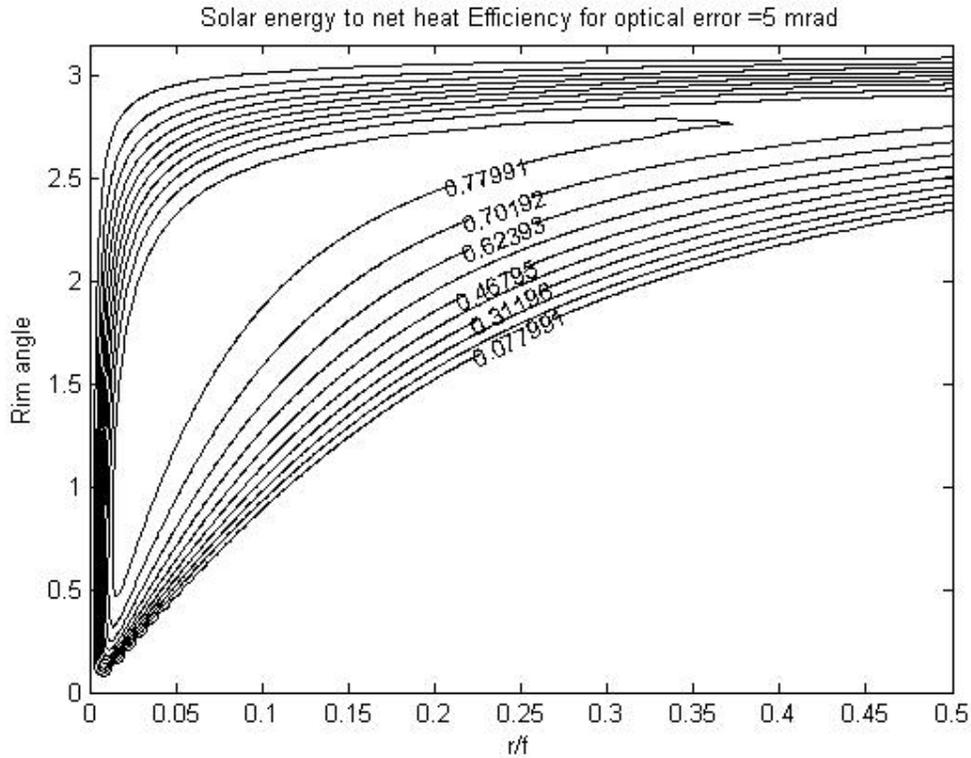

图 8 旋转抛物面球形接收器系统效率与设计参数边缘角和接收器张角的关系。
光学误差为 5mrad，接收器能量损失系统 ξ ＝20.674，反射率和吸收率系数乘积 ρα＝0.8784

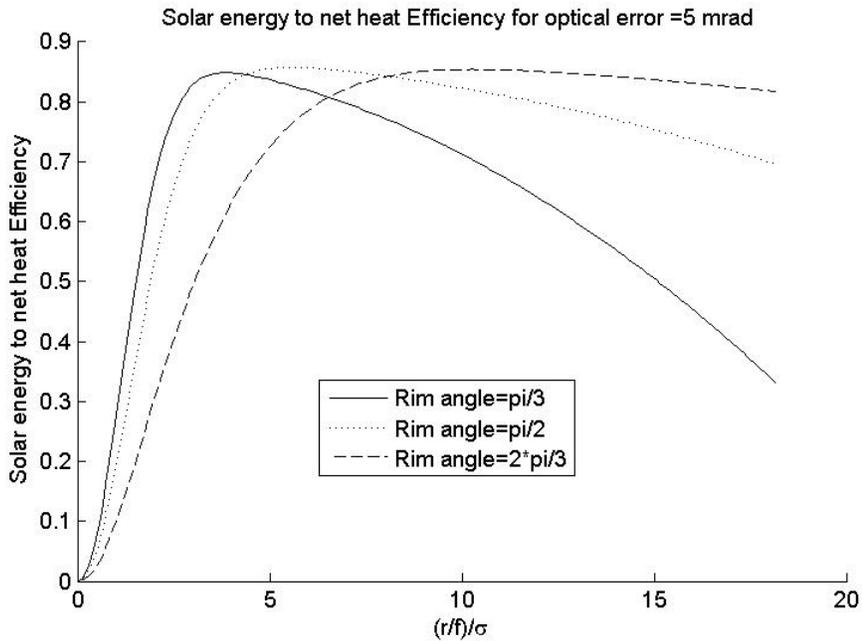

图 9 旋转抛物面球形接收器系统效率与接收器张角对反射光强分布宽度之比的关系。
图 9 是系统光学误差为 5mrad 时，三种边缘角下，旋转抛物面球形接收器系统效率与接收器张角对反射光强分布宽度之比 k 的关系。从图中结果可以看出，边缘角较小时，系统效率随 k 先快速上升，到拦截率接近 100%后，继续增加 k，带来热损增加，效率迅速下降。在边缘较

较大时，效率到达顶点后，计算增加接收器半径，从而增加 k，带来效率缓慢下降，说明大边缘角系统高效率区范围大，更易于制造。

在系统效率最大时，我们有：

$$\partial \eta / \partial \theta_0 = 0$$
$$\partial \eta / \partial \beta_R = 0 \quad (32)$$

将（30）或（31）式代入上式，我们可以得到关于边缘角和接收器张角两个设计参量的二元非线性方程组，它很复杂，没有解析解，但能用数值算法，利用计算机求解，在实际求解时，我们根据（30）或（31）式，采用搜索算法直接求解-η 最小值，从而得到最佳效率点。图 10 给出了计算结果，旋转抛物面球形接收器系统最佳效率与光学误差关系。表 2 是在最佳设计效率下的设计参数。其中光学误差在 5mrad 以下是按（30）式计算，其他按（31）式计算得到。

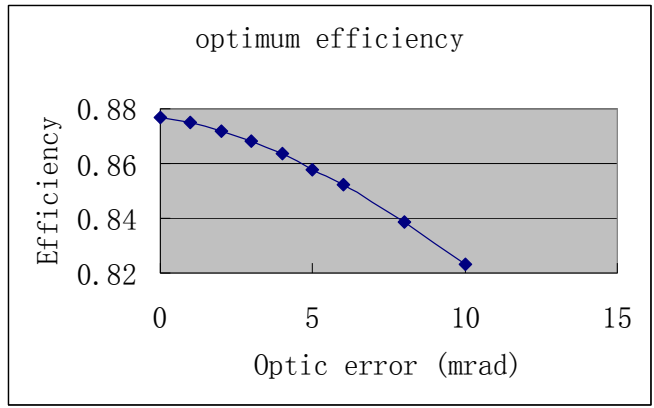

图 10：旋转抛物面球形接收器系统效率最佳效率与光学误差的关系
接收器能量损失系数 ξ＝20.674，反射率和吸收率系数乘积 ρα＝0.8784

| optic error | optimum efficiency | Sigtt | r/f | rim angle |
|---|---|---|---|---|
| 0 | 0.8767 | 2.906 | 0.0097 | 1.5704 |
| 1 | 0.8749 | 2.95 | 0.014 | 1.6209 |
| 2 | 0.8719 | 3.2446 | 0.0189 | 1.6383 |
| 3 | 0.8682 | 3.8568 | 0.0237 | 1.6502 |
| 4 | 0.8636 | 4.6454 | 0.0286 | 1.6613 |
| 5 | 0.8579 | 5.5165 | 0.0336 | 1.6707 |
| 6 | 0.8521 | 6.4302 | 0.0381 | 1.6762 |
| 8 | 0.8386 | 8.3224 | 0.0468 | 1.6871 |
| 10 | 0.8232 | 10.2579 | 0.0552 | 1.6964 |

从计算结果可以看出，由于反射率与吸收率的乘积为 0.8784，在误差低于 5mrad 时，最佳效率非常接近 0.8784，工作温度在 800 度下的接收器热损对效率影响很小，几乎可以忽略不计，而槽式和塔式系统(Reilly and Kolb 2001)的工作温度较低，热损却很大，常常会使系统效率降低 10%左右(Forristall 2003)，从理论上看，这是因为碟式系统聚光比比槽式和塔式系统高，同样反射面下，接收器面积要小得多。此外，碟式系统采用双轴跟踪，余弦因子为 1，也比槽式和塔式系统的余弦因子 0.9(Chen and Li 2003)和 0.8(Yao, Wang et al. 2009)高 10

－20%以上，而系统拦截率也保持在99%以上，因此，采用碟式系统收集能量，总效率要比槽式和塔式系统高20%以上，在效率上是很有优势的。

5、总结：

本文提出的一种计算旋转抛物面聚光光学系统效率的方法，得到了高斯太阳模型和真实太阳模型下，旋转抛物面球形接收器系统拦截率和效率的计算式，表征了系统光学性能与光学误差和主要设计参数，包括接收器张角对反射光强分布宽度之比的关系，评估了高斯模型带来的误差和适用范围。我们给出的计算旋转抛物面球形接收器系统效率的表达式是解析表达式，还给出了表达式中所有参数，可以快速准确计算系统性能。

在此基础上，我们提出了优化系统设计的方法，给出了优化结果。优化结果表明，碟式系统具有较高热效率，主要体现在热损小，余弦因子高，光学效率高等方面，相比塔式和槽式系统，旋转抛物面球形接收器碟式系统在热效率方面具有较大优势。

表 1 计算不同光学误差下，碟式聚光系统反射光线归一化强度分布的多项式拟合参数

| 误差 mrad | $a_0$ | $a_1$ | $a_2$ | $a_3$ | $a_4$ | $a_5$ | $a_6$ | $a_7$ | $a_8$ | $a_9$ | $a_{10}$ |
|---|---|---|---|---|---|---|---|---|---|---|---|
| 0 | 16666.05449 | -104734785 | -2.389846904*10^13 | 1.424615542*10^18 | 3.480460113*10^24 | -1.327605028*10^30 | 2.189947157*10^35 | -1.95048842*10^40 | 9.721333775*10^44 | -2.548063511*10^49 | 2.730480064*10^53 |
| 1 | 16391.78786 | -149034379 | -8.706833464*10^12 | 5.342343505*10^17 | -2.705771566*10^23 | 2.39607337*10^28 | -9.653193212*10^32 | 2.162851262*10^37 | -2.803610917*10^41 | 1.978030042*10^45 | -5.902168901*10^48 |
| 2 | 14547.22215 | -433489856 | -3.540902762*10^12 | 3.851438797*10^17 | -8.335555365*10^21 | 8.515227661*10^25 | -2.748354793*10^29 | -3.231183708*10^33 | 4.20516174*10^37 | -1.961069457*10^41 | 3.467705303*10^44 |
| 3 | 10586.53852 | -319763337 | 4.014695754*10^12 | -1.919080609*10^16 | -1.474745129*10^20 | 3.439157309*10^24 | -3.117834832*10^28 | 1.708254621*10^32 | -5.857097575*10^35 | 1.159640174*10^39 | -1.013007527*10^42 |
| 4 | 7348.160353 | -164372948 | 1.759432956*10^12 | -1.17612074*10^16 | 5.311566706*10^19 | -1.597808668*10^23 | 2.705559179*10^26 | 2.180218427*10^27 | -1.154381591*10^33 | 2.414404609*10^36 | -1.743148139*10^39 |